\documentclass[10pt,notitlepage,a4paper,aps,prd,onecolumn,superscriptaddress,nofootinbib,groupedaddress]{revtex4-1}
\usepackage{array}

\usepackage[dvipsnames]{xcolor}

\usepackage[pdfencoding=auto, psdextra]{hyperref}
\usepackage[T1]{fontenc}
\usepackage[utf8]{inputenc}
\usepackage[english]{babel}
\usepackage{hyperref}
\usepackage[mathscr]{euscript}
\usepackage{caption}
\usepackage{subcaption}
\hypersetup{
colorlinks = true,
linkcolor = blue,
filecolor = magenta,
urlcolor = blue,
pdftitle = {CP3 entanglement},
pdfpagemode = FullScreen,
citecolor = blue}
\usepackage{physics}

\usepackage{caption}

\usepackage{indentfirst}

\usepackage[pdftex]{graphicx}

\usepackage{amsmath}
\allowdisplaybreaks  
\usepackage{amssymb} 
\usepackage{amsfonts} 
\usepackage{enumerate}

\usepackage{fancyhdr}                     

\usepackage{float}                             

\usepackage{amsmath}         
\usepackage{amsfonts}        
\usepackage{amssymb}         
\usepackage{amsthm}          
\usepackage{empheq}          

\usepackage{color}           
\usepackage{ragged2e}        
\usepackage{titlesec}        

\usepackage{tipa}            

\usepackage[utf8]{inputenc} 
\allowdisplaybreaks          

\usepackage{enumitem}        

\usepackage{tikz}            
\usetikzlibrary{shapes.geometric,arrows.meta,arrows,positioning}  
\usetikzlibrary{tikzmark}   

\titleformat{\paragraph}
  {\normalfont\normalsize\bfseries}{\theparagraph}{1em}{} 
\titlespacing*{\paragraph}
  {0pt}{3.25ex plus 1ex minus .2ex}{1.5ex plus .2ex} 

\DeclareUnicodeCharacter{202F}{\,}
\usepackage[utf8]{inputenc}
\begin{document}
\title{Exact Classicalization of $N$-Level Quantum Systems Interacting with a Bath: Theory and Applications}

\author{Daniel Martínez-Gil}
\email{daniel.martinez@ua.es}
\affiliation{Fundacion Humanismo y Ciencia, Guzmán el Bueno, 66, 28015 Madrid, Spain.}
\affiliation{Departamento de F\'{\i}sica Aplicada, Universidad de Alicante, Campus de San Vicente del Raspeig, E-03690 Alicante, Spain.}

\author{Pedro Bargueño}
\email{pedro.bargueno@ua.es}
\affiliation{Departamento de F\'{\i}sica Aplicada, Universidad de Alicante, Campus de San Vicente del Raspeig, E-03690 Alicante, Spain.}

\author{Salvador Miret-Artés}
\email{s.miret@iff.csic.es}
\affiliation{Instituto de Física Fundamental, Consejo Superior de Investigaciones Científicas, Serrano 123, 28006, Madrid, Spain}

\begin{abstract}
 In this manuscript, starting from a five-step algorithmic procedure for exactly classicalizing the dynamics of $N$-level quantum systems, we incorporate a classical bath of harmonic oscillators to model environmental interactions.
Using the geometry of complex projective spaces $\mathbb{CP}^{N-1}$ and a Langevin formalism, we
obtain $N-1$ Hamilton’s equations which encode both the quantum system and the bath degrees of freedom, representing a generalization of the Caldeira-Legget model in a complex projective space.
We demonstrate the efficacy of the method by applying it to two paradigmatic systems: a two-qubit system in $\mathbb{CP}^3$ under entangling interactions, reproducing quantum observables such as state populations, quaternionic population differences and concurrence, and the seven-state Fenna-Matthews-Olson (FMO) complex in $\mathbb{CP}^6$, reproducing state populations in the picosecond timescale.
    
\end{abstract}

\maketitle

\section{Introduction}
Open quantum systems, characterized by their interactions with external environments, are essential to a wide range of physical phenomena, from the decoherence processes that challenge quantum technologies \cite{QT1, QT2, QT3,QT4,QT5,QT6} to the remarkably efficient energy transfer observed in biological systems such as the Fenna-Matthews-Olson (FMO) complex \cite{FennaMatthews1975, AdolphsRenger2006}. These systems, among others, imply significant computational challenges, as traditional quantum simulation techniques often demand substantial resources or use approximations that may compromise accuracy, making classical interpretations widely researched in this context \cite{strocchi, Meyermiller, StockThoss, Tully1990,CottonMiller, Richardson1, Richardson2}.
In this article, we introduce an exact classicalization technique for systems (originally developed for isolated quantum $N$-level systems \cite{dani}), extending it to open quantum systems by coupling it to a classical bath of harmonic oscillators, maintaining the mathematical rigor of our previous work. 

The main idea of the method here employed lies in the geometric structure of complex projective spaces, denoted as $\mathbb{CP}^{N-1}$, which serves as the natural state space for an $N$-level quantum system.  Mathematically, $\mathbb{CP}^{N-1}$ can be classified as a Kähler manifold \cite{Moroianu_2007, Bellmann, Nakahara}, having a complex structure \cite{chern-1979, Griffiths, harris-1992} and both a Fubini-Study metric (defining distances) \cite{fubini, study} and a compatible symplectic form (enabling Hamiltonian dynamics) \cite{arnold, Lee}. This formalism provides a method
 for transforming quantum evolution into a classical-like Hamiltonian system, providing a geometric perspective of classicalization \cite{Kibble, GIBBONS1992147, Ashtekar1999, BRODY200119}. Specifically, the methodology is based on a five-step method previously introduced for isolated quantum systems \cite{dani}, which will be briefly summarized in the present manuscript.  This process reduces the quantum problem to a set of $N-1$ Hamilton equations, corresponding to the $N-1$ complex dimensions of $\mathbb{CP}^{N-1}$, while preserving exactly the quantum behavior.

In this work, we extend this geometric approach to open quantum systems by coupling the classicalized system to a classical bath of harmonic oscillators, which models usual quantities in this context, such as
the environment’s dissipative and noisy effects \cite{weiss,Schlosshauer2007} as it will be shown. 
The power of this framework is demonstrated through two illustrative applications. In Section \ref{entanglementbath}, we examine the dynamics of two entangled qubits subjected to environmental effects via the classical bath. We explore how dissipation affects the system, offering insights into entanglement under more realistic conditions using the classical formalism. In Section \ref{fmo}, we apply the method to the seven-state FMO complex, a photosynthetic system where quantum coherence enhances energy transfer efficiency.  

The article is structured as follows: Section \ref{method} introduces the five-step classicalization method for $N$-state systems, emphasizing its geometric background; Section \ref{systembath} details the coupling to the classical bath and the resulting general equations of motion; Section \ref{entanglementbath} analyzes the entangled qubit system under environmental influence; and Section \ref{fmo} explores the FMO complex’s dynamics. Section \ref{conclusion} summarizes key findings and outlines the conclusions of the present work.

\section{Classical N-state system method}\label{method}
In a previous work, we presented an exact five-step method to classicalize any $N$-level quantum system \cite{dani}, based on the geometric properties of its state space, the complex projective space $\mathbb{CP}^{N-1}$ \cite{book_geometry_quantum}. This space is a Kähler manifold \cite{Moroianu_2007, Bellmann, Nakahara}, meaning it is equipped with a symplectic structure \cite{arnold, Lee} and a compatible metric, the Fubini-Study metric. This geometric structure allows us to reformulate quantum dynamics in terms of Hamiltonian dynamics on $\mathbb{CP}^{N-1}$, transforming quantum equations into classical trajectories without loss of information. Below, we outline the five steps of the method:

\begin{enumerate}
    \item \textbf{Definition of coordinates in $\mathbb{CP}^{N-1}$:} Complex coordinates $x^j$ are introduced in $\mathbb{CP}^{N-1}$ using a reference state, providing an explicit parametrization of the quantum state space.

The wave function for a system of $N$ states can be expressed as
\begin{equation}
\ket{\psi} = \sum_{i=0}^{N-1} a^i \ket{\phi_i},
\end{equation}
where $a^i$ are complex coefficients and $\ket{\phi_i}$ are the basis states. 
To define coordinates $x^j$ in $\mathbb{CP}^{N-1}$, we select a non-zero coefficient ($a^{N-1}$), expressing the coordinates as
\begin{equation}
x^j = \frac{a^j}{a^{N-1}} \quad \text{for} \quad j = 0, 1, \ldots, N-2,
\end{equation}
obtaining $N-1$ independent complex coordinates.

    \item \textbf{Expression of the wave function:} The quantum wave function is rewritten in $\mathbb{CP}^{N-1}$ coordinates as
    \begin{equation}
\ket{\psi} = \frac{1}{\sqrt{\mathcal{N}}} \left( \sum_{i=0}^{N-2} x^i \ket{\phi_i} + \ket{\phi_{N-1}} \right),
\end{equation}
    incorporating a normalization factor
    \begin{equation}
        \mathcal{N} = 1 + \sum_{i=0}^{N-2} |x^i|^2,
    \end{equation}
    that ensures the probabilistic consistency of the state.

    \item \textbf{Calculation of the classical Hamiltonian:} The classical Hamiltonian is obtained as the expectation value of the quantum Hamiltonian projected onto $\mathbb{CP}^{N-1}$ wave function as
\begin{equation}
     H =    \langle \psi | \hat{H} | \psi \rangle.
\end{equation}
    
    \item \textbf{Obtaining the symplectic form:} From the Kähler potential
    \begin{equation}
K = \log \left( \mathcal{N} \right),
\end{equation}
the symplectic form $\omega$ is derived, defining the Poisson structure and enabling the formulation of classical dynamics. In particular, the inverse symplectic form matrix,
\begin{equation}
    \omega^{jk} = \left(i \frac{\partial^2 K}{\partial x^j \partial \bar{x}^k}\right)^{-1}
    =
    -i \mathcal{N} \left( \delta^{j k} + x^j \bar{x}^k \right),
\end{equation}
is used, where $\delta^{jk}$ is the  Kronecker delta, and $\bar{x}^k$ represent the complex conjugate of $x^k$.

    \item \textbf{Hamilton's equations:} Using the inverse symplectic form, the Poisson bracket using pairs of canonically conjugate variables $(q^i,p^i)$, is defined as
\begin{equation}\label{poisson}
    \{f,g\} = \sum_{j,k = 0}^{N-2}\omega^{jk} \left(\frac{\partial f}{\partial q^j}\frac{\partial g}{\partial p^k}-\frac{\partial f}{\partial p^j}\frac{\partial g}{\partial q^k}\right).
\end{equation}

This enables us to define  Hamilton's equations governing the time evolution, which are obtained as
    \begin{equation}
\dot{x}^j = \{x^j, H_S\} = \sum_{k=0}^{N-2} -i \mathcal{N} (\delta^{j k} + x^j \bar{x}^k)\frac{\partial H_S}{\partial \bar{x}^k}.
\end{equation}
\end{enumerate}

We would like to remark that this method is exact, preserving the full quantum dynamics, including features like entanglement, within a unified geometric framework. Furthermore, it reduces the original $N$ quantum equations to $N-1$ classical equations, enabling the simulation of quantum behavior in a lower-dimensional classical formalism.

\section{System plus environment approach}\label{systembath}

In this section, we will study the system of interest as an open quantum system, considering the previously presented method.
 The system plus bath formalism provides a robust framework for this purpose, where the system is coupled to an external bath that introduces effects such as decoherence and dissipation. Typically, the bath is modeled as a collection of harmonic oscillators  \cite{caldeiraleggett1, Caldeiralegget2, CaldeiraLegget3,RevModPhys.59.1,THORWART2004333, PhysRevA.68.034301, PhysRevB.71.035318}, which interact with the system to usually produce a decoherence effect on it.  In this section, we represent the total system plus bath formalism in a classical framework. Specifically, we use the classical dynamics formulated on the complex projective space $\mathbb{CP}^{N-1}$, as developed previously, and couple the $\mathbb{CP}^{N-1}$ coordinates to a bath of classical harmonic oscillators. In the following, we detail the methodology for this coupling and how to obtain Hamilton's equations for the total system plus bath, with Hamiltonian
 \begin{equation}
    H_T = H_S + H_B+H_I,
\end{equation}
where $H_S, H_{B}$ and $H_{I}$ stand for the Hamiltonian of the system, bath and interaction, respectively.

We consider the Hamiltonian of certain $N$-level system in terms of the coordinates of $\mathbb{CP}^{N-1}$ as
\begin{equation}
    H_S = H_S (x^j, \bar{x}^j) \in \mathbb{CP}^{N-1},
\end{equation}
and the Hamiltonian of classical harmonic oscillators as
\begin{equation}
    H_B = \sum_i^n\left( \frac{p_i^2}{2m_i} + \frac{m_i}{2}\omega_i^2 q_i^2\right) \in \mathbb{R}^{2n},
\end{equation}
where $q_i, p_i, m_i, \omega_i$, represent the harmonic oscillator's position, momentum, mass and angular frequency, respectively, and $n$ is the number of harmonic oscillators in the bath.

In order to preserve the reality of the Hamiltonian, and to consider the real and the imaginary part of the coordinates, we choose to couple the bath to $\abs{x^j}^2$. Therefore, we will consider the interacting  Hamiltonian as
\begin{equation}\label{int_Ham}
    H_I = -\sum_{j = 0}^{N-1} \abs{x^j}^2 \sum_i^n C_{ij} q_i + \sum_{j = 0}^{N-1}  \abs{x^j}^4  \sum_i^n \frac{C_{ij}^2}{2m_i \omega_i^2}, 
\end{equation}
where we have considered the coupling between all $\abs{x^j}^2$ and the positions $q^j$, and $C_{ij}$ represents the coupling strength between them. Regarding the last term of the Eq. \eqref{int_Ham}, some comments are in order.

In the study of open quantum systems, it is usual to consider the interaction Hamiltonian with a counterterm \cite{weiss}, represented in our model by the second term in Eq. \eqref{int_Ham}, to address the renormalization effects induced by the coupling between the system and the bath. This counterterm is primarily employed to compensate for the shift in the effective potential, ensuring that the intrinsic dynamics of the system, described by $ H_S $, remain unperturbed.  Without this adjustment, the coupling term would introduce an artificial modification to the system's energy, as the bath oscillators adjust to the new equilibrium position 
\begin{equation}
    q_i(eq) = \sum_{j=0}^{N-1}\frac{\abs{x^j}^2C_{ij}}{m_i \omega_i^2}
\end{equation}
leading a the potential shift equal to
\begin{equation}
    \nabla V= \sum_{j = 0}^{N-1}  \abs{x^j}^4  \sum_i^n \frac{C_{ij}^2}{2m_i \omega_i^2},
\end{equation}
which is the renormalization term in the Hamiltonian. By including the counterterm, we preserve the physical integrity of the system's Hamiltonian, aligning with the framework established in models such as Caldeira-Leggett \cite{caldeiraleggett1,Caldeiralegget2,CaldeiraLegget3}, where such corrections are essential for consistent dissipative dynamics.

The challenge in coupling the classical dynamics of the system on $\mathbb{CP}^{N-1}$ with a bath of harmonic oscillators in $\mathbb{R}^{2n}$ lies in their distinct symplectic structures, which lead to different forms of Hamilton's equations. The space $\mathbb{CP}^{N-1}$ employs the symplectic form introduced earlier in this work (the interested reader can read \cite{dani} for more detail), while $\mathbb{R}^{2n}$ uses the canonical symplectic form, resulting in the standard Hamilton's equations for a phase space of positions and momenta.

This issue is resolved by defining the total phase space as $\Gamma = \mathbb{CP}^{N-1} \times \mathbb{R}^{2n}$. The total phase space $\Gamma$ is a product of two symplectic manifolds: $\mathbb{CP}^{N-1}$, equipped with the Kähler symplectic form $\omega_{\mathbb{CP}^{N-1}}$, and $\mathbb{R}^{2n}$, equipped with the canonical symplectic form $\omega_{\mathbb{R}^{2n}}$. In a direct product of symplectic manifolds, the total symplectic form is the direct sum $\omega_\Gamma = \omega_{\mathbb{CP}^{N-1}} + \omega_{\mathbb{R}^{2n}}$, reflecting the independence of the coordinates $x^j, \bar{x}^k$ and $q^i, p^i$. The associated Poisson bracket is defined as 
\begin{equation}
    \{f, g\}_\Gamma = \sum_{i,j} \omega_\Gamma^{ij} \frac{\partial f}{\partial r^i} \frac{\partial g}{\partial r^j},
\end{equation}
where the inverse symplectic tensor $\omega_\Gamma^{ij}$ is block-diagonal, and $r^i, r^j$ can be any of the coordinates of $\mathbb{CP}^{N-1}$ or $\mathbb{R}^{2n}$. Consequently, the cross-terms between $\mathbb{CP}^{N-1}$ and $\mathbb{R}^{2n}$ vanish, yielding $\{f, g\}_\Gamma$ as the direct sum of the individual brackets, as
\begin{equation}\label{totalPoisson}
    \{f, g\}_\Gamma = \{f, g\}_{\mathbb{CP}^{N-1}} + \{f, g\}_{\mathbb{R}^{2n}},
\end{equation}
where the Poisson brackets for each subspace are given by:
\begin{align}
    \{f, g\}_{\mathbb{CP}^{N-1}} &= \sum_{j,k=0}^{N-2} \omega^{j k} \left( \frac{\partial f}{\partial x^j} \frac{\partial g}{\partial \bar{x}^k} - \frac{\partial f}{\partial \bar{x}^k} \frac{\partial g}{\partial x^j} \right), \\
    \{f, g\}_{\mathbb{R}^{2n}} &= \sum_{i=1}^n \left( \frac{\partial f}{\partial q^i} \frac{\partial g}{\partial p^i} - \frac{\partial f}{\partial p^i} \frac{\partial g}{\partial q^i} \right).
\end{align}

Hamilton's equations for the total system are thus defined in terms of the total Hamiltonian $H_T$. 
Since the system Hamiltonian and the bath Hamiltonian do not depend on each other's coordinates, certain derivatives vanish, leading to the following equations of motion:
\begin{align}
    \dot{x}^j &= \{ x^j, H_T \}_\Gamma = \{ x^j, H_S \}_{\mathbb{CP}^{N-1}} + \{ x^j, H_I \}_{\mathbb{CP}^{N-1}}, \\
    \dot{q}^i &= \{ q^i, H_T \}_\Gamma = \{ q^i, H_B \}_{\mathbb{R}^{2n}} + \{ q^i, H_I \}_{\mathbb{R}^{2n}}, \\
    \dot{p}^i &= \{ p^i, H_T \}_\Gamma = \{ p^i, H_B \}_{\mathbb{R}^{2n}} + \{ p^i, H_I \}_{\mathbb{R}^{2n}}.
\end{align}

For the system coordinates $x^j$ on $\mathbb{CP}^{N-1}$, the equations of motion take the explicit form
\begin{equation}
    \dot{x}^j = \sum_{k=0}^{N-2} -i \mathcal{N} \left( \delta^{jk} + x^j \bar{x}^k \right) \left( \frac{\partial H_S}{\partial \bar{x}^k} + \frac{\partial H_I}{\partial \bar{x}^k} \right),
\end{equation}
where $\mathcal{N} = 1 + \sum_{i=0}^{N-2} |x^i|^2$ is the normalization factor. These equations govern the dynamics of the system coordinates, incorporating the influence of the bath through the interaction term $H_I$.

At this point, we will analytically derive how the interaction affects the system. Following Eq. \eqref{int_Ham}, the interaction Hamiltonian derivative can be expressed as
\begin{align}
    \frac{\partial H_I}{\partial \bar{x}^j} &= -x^j \sum_i C_{ji} q_i+ x^j\abs{x^j}^2 \sum_i \frac{C_{ij}^2}{m_i\omega_i^2},
\end{align}
which represents a generalization of the Caldeira-Leggett model \cite{caldeiraleggett1,Caldeiralegget2, CaldeiraLegget3} in the projective space $\mathbb{CP}^N$.

On the other hand, the bath, modeled as a set of harmonic oscillators, follows the dynamics
\begin{align}
    \dot{p}_i + m_i \omega_i^2 q_i &= C_{0i} |x^0|^2 + C_{1i} |x^1|^2 + C_{2i} |x^2|^2, \\
    \dot{p}_i &= m_i \ddot{q}_i,
\end{align}
where $p_i$ and $q_i$ are the momentum and position of the $i$-th bath oscillator, and $|x^j|^2 = x^j \bar{x}^j$ reflects the system’s influence on the bath.

One great advantage of this model lies in the fact that the environmental degrees of freedom can be exactly eliminated from the
equations of motion by using its general solution, based on the Langevin formalism \cite{weiss,PedroCPL2011, Pedro2012}. Using this kind of formalism, the interaction Hamiltonian derivatives now read as
\begin{align}\label{intHamdamping}
    \frac{\partial H_I}{\partial \bar{x}^j} &= -x^j \left(\xi_j(t) - \int_0^t \gamma_j(t-s)\frac{d}{dt}\abs{x^j}^2 ds\right), 
\end{align}
where the stochastic noise $\xi_j(t)$ is expressed as
\begin{equation}\label{noise}
    \xi_j(t) = \sum_i^n C_{ij}\left[\left(q_i(0)-\abs{x^j(0)}^2\frac{C_{ij}}{m_i\omega_i^2}\right)\cos(\omega_it)+\frac{p_i(0)}{m_i\omega_i}\sin(\omega_it)\right],
\end{equation}
and the damping kernel $\gamma_j(t)$ is given by
\begin{equation}\label{damping}
    \gamma_j(t) = \sum_i^n \frac{C_{ij}^2}{m_i\omega_i^2}\cos(\omega_i t).
\end{equation}

As we can see, equation \eqref{intHamdamping} represents a generalization of the Caldeira-Leggett model within the Langevin formalism in $\mathbb{CP}^N$ coordinates, where the stochastic noise \eqref{noise} and the damping kernel \eqref{damping} take roughly their usual forms.

In order to work with a converged number of harmonic oscillators in the bath, we adopt an Ohmic spectral density, $J(\omega) = \frac{2\delta}{K} \gamma \omega$, where $\delta$, $K$, and $\gamma$ are constants, reflecting a linear frequency dependence (it has to be noted that the spectral density corresponds to one considering a high-frequency Drude cutoff $J(\omega) = \frac{2\delta}{K} \gamma \omega e^{-\omega/\omega_c}$). In the Markovian approximation, where the bath’s correlation time is short compared to the system’s dynamics, the damping kernel simplifies to

\begin{equation}
\gamma_j(t) \approx 2 \gamma_j \delta(t),
\end{equation}
 indicating instantaneous dissipation with negligible memory effects. Moreover, as the noise is a stochastic variable, in the following sections we will suppose an average over different simulations. This permits us to maintain the same parameters for the system, leading this random force to vanish, and obtaining pure dissipative dynamics, our results being temperature-independent. For simplicity, we will refer as $\mathtt{x}^j$ to these noise-averaged coordinates.
 Consequently, the interaction Hamiltonian derivatives reduce to

\begin{equation}\label{partial_HI}
\frac{\partial H_I}{\partial \bar{\mathtt{x}}^j} = 2\mathtt{x}^j \gamma_j \frac{d}{dt} |\mathtt{x}^j|^2 .
\end{equation}

Therefore, as a conclusion of this section, we can state that for a quantum system with $N$ states coupled to a bath with $2n$ degrees of freedom, the total number of equations to be solved would be $N + 2n$. In contrast, our model enables the classicalization of that system, considering also the bath’s influence, by requiring only $N-1$ differential equations. This significant reduction in complexity highlights the efficiency of the proposed approach for studying dissipative quantum dynamics.

\section{Entanglement+bath}\label{entanglementbath}
Having established the general classicalized approach for the system plus bath model, we will now apply it to a specific case by examining two entangled qubits coupled to a bath of harmonic oscillators. Although the quantum Hamiltonian for the two-entangled-qubit system can be expressed in terms of 15 different matrices, we will employ a subset of this basis as in our previous work \cite{dani}. This subset captures the general case, including both entangling and non-entangling terms in the Hamiltonian, and is given by
\begin{equation}\label{Hamcuantico}
    \hat{H}_S = C_1 (\hat{\sigma}_z \otimes \hat{I}) + C_2 (\hat{\sigma}_x \otimes \hat{I}) + C_3 (\hat{\sigma}_y \otimes \hat{I}) + C_4 (\hat{\sigma}_y \otimes \hat{\sigma}_y) + C_5 (\hat{\sigma}_x \otimes \hat{\sigma}_y),
\end{equation}

where $\hat{\sigma}_i$ represent the Pauli matrices, and $\hat{I}$ denotes the $4 \times 4$ identity matrix. We also express the quantum state as
\begin{equation}\label{wavefunction}
    \ket{\psi} = a \ket{00}+b\ket{10}+c\ket{01}+d\ket{11}.
\end{equation}

By using the coordinates of  $\ket{\psi}$ on $\mathbb{CP}^3$, the classicalized Hamiltonian is expressed as

\begin{align}\label{systemHamentangled}
& H_S = \bra{\psi} \hat{H}_S \ket{\psi} \nonumber = \frac{1}{\mathcal{N}}\big[ 
C_1(\abs{x^0}^2+\abs{x^1}^2-\abs{x^2}^2-1) 
+ C_2(x^2\Bar{x}^0+\Bar{x}^1+\Bar{x}^2x^0+x^1)+ \nonumber \\
&\quad + i C_3(-x^2\Bar{x}^0-\Bar{x}^1+\Bar{x}^2x^0+x^1)
+ C_4(-\Bar{x}^0+x^2\Bar{x}^1+x^1\Bar{x}^2-x^0) \quad + iC_5(-\Bar{x}^0+x^2\Bar{x}^1-x^1\Bar{x}^2+x^0) 
\big].
\end{align}

For this specific case of two entangled qubits interacting with a bath, the equations of motion for the noise-averaged system coordinates $\mathtt{x}^0$, $\mathtt{x}^1$, and $\mathtt{x}^2$ in $\mathbb{CP}^3$ are given by

\begin{align}
    \dot{\mathtt{x}}^0 &= -i \mathcal{N} \left[ (1 + \mathtt{x}^0 \bar{\mathtt{x}}^0) \left( \frac{\partial H_S}{\partial \bar{\mathtt{x}}^0} + \frac{\partial H_I}{\partial \bar{\mathtt{x}}^0} \right) + \mathtt{x}^0 \bar{\mathtt{x}}^1 \left( \frac{\partial H_S}{\partial \bar{\mathtt{x}}^1} + \frac{\partial H_I}{\partial \bar{\mathtt{x}}^1} \right) + \mathtt{x}^0 \bar{\mathtt{x}}^2 \left( \frac{\partial H_S}{\partial \bar{\mathtt{x}}^2} + \frac{\partial H_I}{\partial \bar{\mathtt{x}}^2} \right) \right], \\
    \dot{\mathtt{x}}^1 &= -i \mathcal{N} \left[ \mathtt{x}^1 \bar{\mathtt{x}}^0 \left( \frac{\partial H_S}{\partial \bar{\mathtt{x}}^0} + \frac{\partial H_I}{\partial \bar{\mathtt{x}}^0} \right) + (1 + \mathtt{x}^1 \bar{\mathtt{x}}^1) \left( \frac{\partial H_S}{\partial \bar{\mathtt{x}}^1} + \frac{\partial H_I}{\partial \bar{\mathtt{x}}^1} \right) + \mathtt{x}^1 \bar{\mathtt{x}}^2 \left( \frac{\partial H_S}{\partial \bar{\mathtt{x}}^2} + \frac{\partial H_I}{\partial \bar{\mathtt{x}}^2} \right) \right], \\
    \dot{\mathtt{x}}^2 &= -i \mathcal{N} \left[ \mathtt{x}^2 \bar{\mathtt{x}}^0 \left( \frac{\partial H_S}{\partial \bar{\mathtt{x}}^0} + \frac{\partial H_I}{\partial \bar{\mathtt{x}}^0} \right) + \mathtt{x}^2 \bar{\mathtt{x}}^1 \left( \frac{\partial H_S}{\partial \bar{\mathtt{x}}^1} + \frac{\partial H_I}{\partial \bar{\mathtt{x}}^1} \right) + (1 + \mathtt{x}^2 \bar{\mathtt{x}}^2) \left( \frac{\partial H_S}{\partial \bar{\mathtt{x}}^2} + \frac{\partial H_I}{\partial \bar{\mathtt{x}}^2} \right) \right],
\end{align}

where $\mathcal{N} = 1 + |\mathtt{x}^0|^2 + |\mathtt{x}^1|^2 + |\mathtt{x}^2|^2$ is the normalization factor, preserving the projective nature of the coordinates.

In this particular scenario, the partial derivatives of the system Hamiltonian are

\begin{align}
    \frac{\partial H_S}{\partial \bar{\mathtt{x}}^0} &= \frac{(C_1 \mathtt{x}^0 + C_2 \mathtt{x}^2 - i C_3 \mathtt{x}^2 - C_4 - i C_5) \mathcal{N} - D \mathtt{x}^0}{\mathcal{N}^2}, \label{x0} \\
    \frac{\partial H_S}{\partial \bar{\mathtt{x}}^1} &= \frac{(C_1 \mathtt{x}^1 + C_2 - i C_3 + C_4 \mathtt{x}^2 + i C_5 \mathtt{x}^2) \mathcal{N} - D \mathtt{x}^1}{\mathcal{N}^2}, \label{x1} \\
    \frac{\partial H_S}{\partial \bar{\mathtt{x}}^2} &= \frac{(-C_1 \mathtt{x}^2 + C_2 \mathtt{x}^0 + i C_3 \mathtt{x}^0 + C_4 \mathtt{x}^1 - i C_5 \mathtt{x}^1) \mathcal{N} - D \mathtt{x}^2}{\mathcal{N}^2}, \label{x2}
\end{align}

where the system Hamiltonian \eqref{systemHamentangled} has been expressed as $H_S = \frac{D}{\mathcal{N}} $. Additionally, the derivatives of the interaction Hamiltonian in the noise-averaged coordinates are
\begin{align}
    \frac{\partial H_I}{\partial \bar{\mathtt{x}}^0} &= 2 \mathtt{x}^0 \gamma_0 \frac{d}{dt} |\mathtt{x}^0|^2, \\
    \frac{\partial H_I}{\partial \bar{\mathtt{x}}^1} &= 2 \mathtt{x}^1 \gamma_1 \frac{d}{dt} |\mathtt{x}^1|^2, \\
    \frac{\partial H_I}{\partial \bar{\mathtt{x}}^2} &= 2 \mathtt{x}^2 \gamma_2 \frac{d}{dt} |\mathtt{x}^2|^2.
\end{align}

As we can see, in contrast to the standard Caldeira-Leggett model, three distinct damping rates appear, one for each coordinate in the complex projective space $\mathbb{CP}^3$.

We want to remark that the environmental degrees of freedom are represented in the derivatives of the interaction Hamiltonian, the three equations \eqref{x0}-\eqref{x2} being enough to analyze the dynamics of the whole system plus bath. With the formalism now established, we proceed to discuss the results.

\subsection{Results and discussion}

In this section, we aim to show how the classical-like two entangled qubits system behaves under its coupling to a bath of harmonic oscillators, considering $\gamma_1 = \gamma_2 = \gamma_3$ in all simulations. Therefore, we will change the value of $\gamma_i$, considering both the cases with entangling ($C_4 = C_5 = 1$) and non-entangling ($C_4 = C_5 = 0$) terms in the Hamiltonian. We will also consider the following parameters at initial times $a(0) = \sqrt{0.4}$ $, b(0) = \sqrt{0.4},$ $ c(0) = 0,$ $ d(0) = \sqrt{0.2}$.

First of all, we present the quantum populations $\abs{a}^2$, $\abs{b}^2$, $\abs{c}^2$ as functions of time, which can be expressed in terms of the noise-averaged coordinates on $\mathbb{CP}^3$ as
\begin{align}
    \abs{a}^2 &= \frac{\abs{\mathtt{x}^0}^2}{\mathcal{N}}, &
    \abs{b}^2 &= \frac{\abs{\mathtt{x}^1}^2}{\mathcal{N}}, &
    \abs{c}^2 &= \frac{\abs{\mathtt{x}^2}^2}{\mathcal{N}}.
\end{align}

It should be noted that $\abs{d}^2$ is not included in the figures to avoid visual overload. Moreover, this quantity depends solely on the chosen normalization, as $\abs{d}^2 = \frac{1}{\mathcal{N}}$.

\begin{figure}[H]
    \centering
    \begin{subfigure}[b]{0.34\textwidth}
        \includegraphics[width=\textwidth]{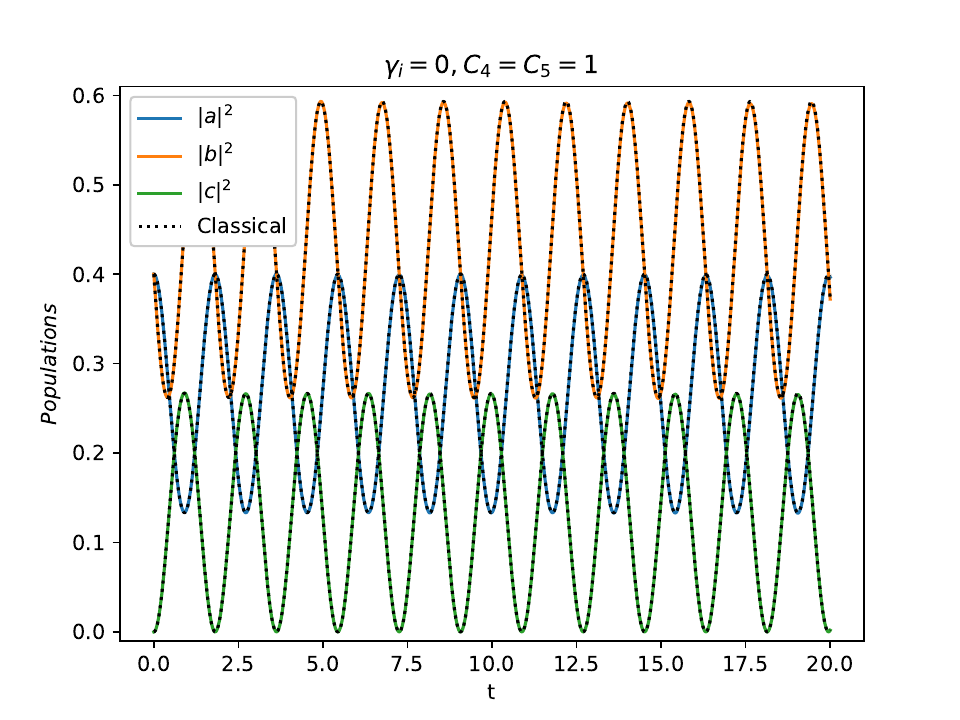}
    \end{subfigure}
    \hspace{0.02\textwidth}
    \begin{subfigure}[b]{0.34\textwidth}
        \includegraphics[width=\textwidth]{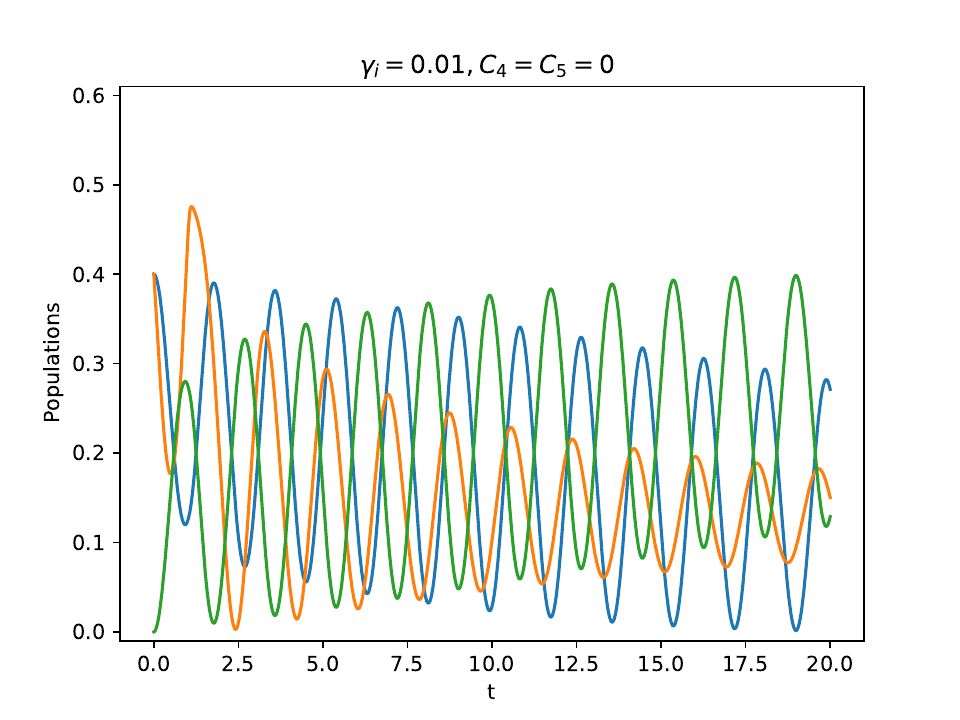}
    \end{subfigure}

    \vspace{0.5em} 

    \begin{subfigure}[b]{0.34\textwidth}
        \includegraphics[width=\textwidth]{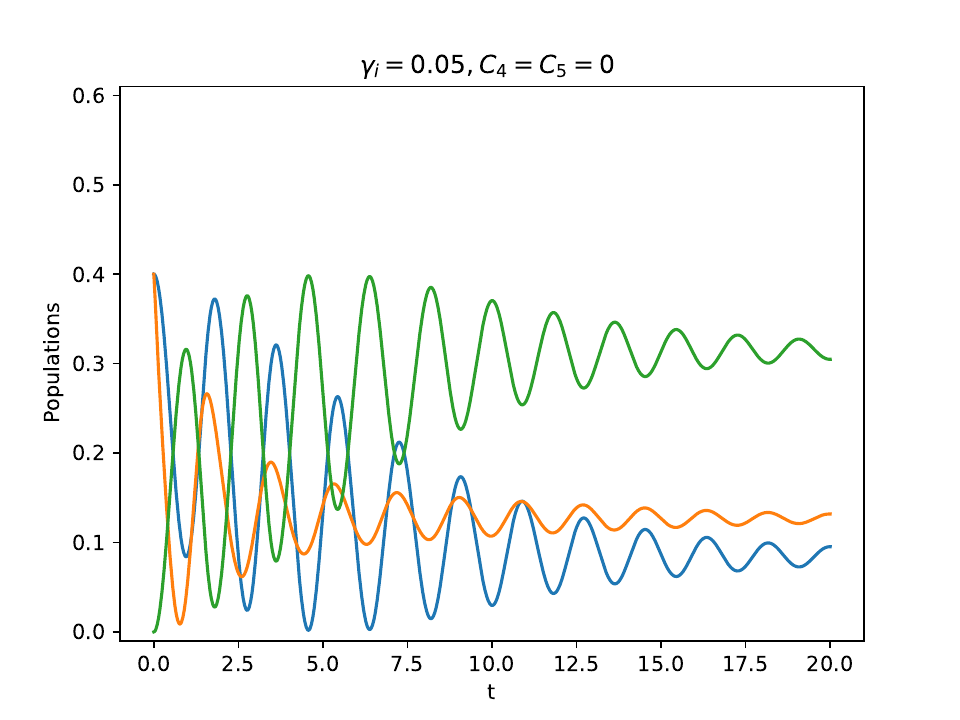}
    \end{subfigure}
    \hspace{0.02\textwidth}
    \begin{subfigure}[b]{0.34\textwidth}
        \includegraphics[width=\textwidth]{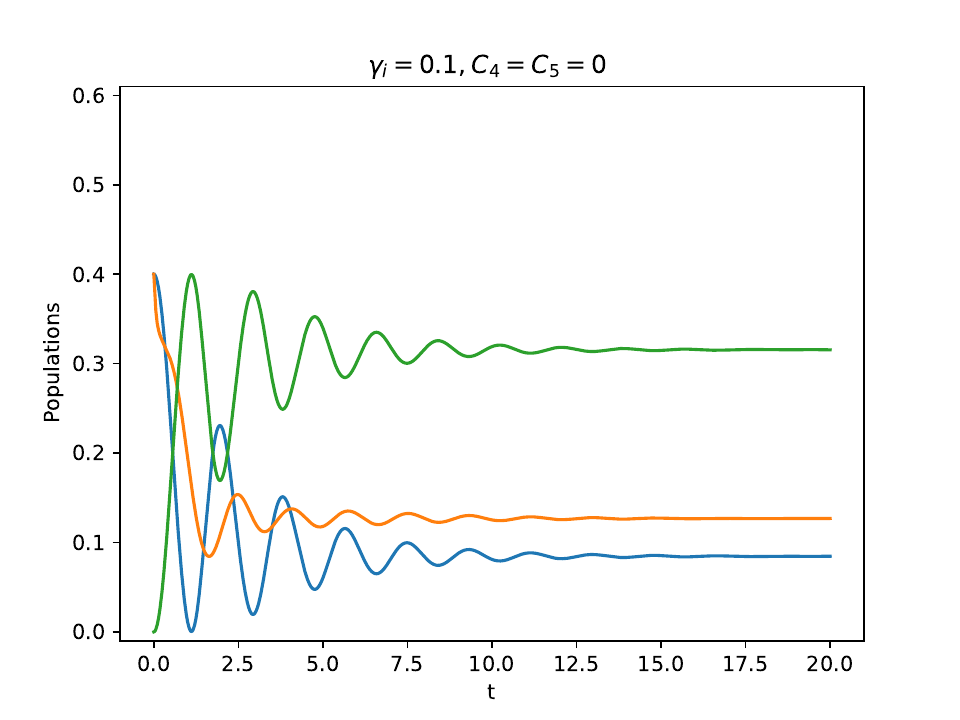}
    \end{subfigure}

    \caption{Time dependent populations varying $\gamma_i$, considering $C_4 = C_5 = 0$. See text for details.}
    \label{populations1}
\end{figure}

As shown in Fig. \ref{populations1}, we set $C_4 = C_5 = 0$, observing a pronounced damping effect that intensifies as $\gamma_i$ increases, resulting in the characteristic behavior of a damped harmonic oscillator. This damping, analogous to the frictional effect in a classical system, reflects the energy dissipation from the quantum system to the environment, leading to a progressive suppression of oscillations in the quantum state populations and showing an environment-induced decoherence effect. In the top-left panel, the colored lines represent the solution of the Schrödinger equation for the populations, while the black lines correspond to the Hamilton solution expressed in $\mathbb{CP}^3$ coordinates, showing the exact quantum-classical correspondence for the system. Although both approaches yield identical results, it should be noted that the remaining panels employ the classical framework to model the system.

\begin{figure}[H]
    \centering
    \begin{subfigure}[b]{0.34\textwidth}
        \includegraphics[width=\textwidth]{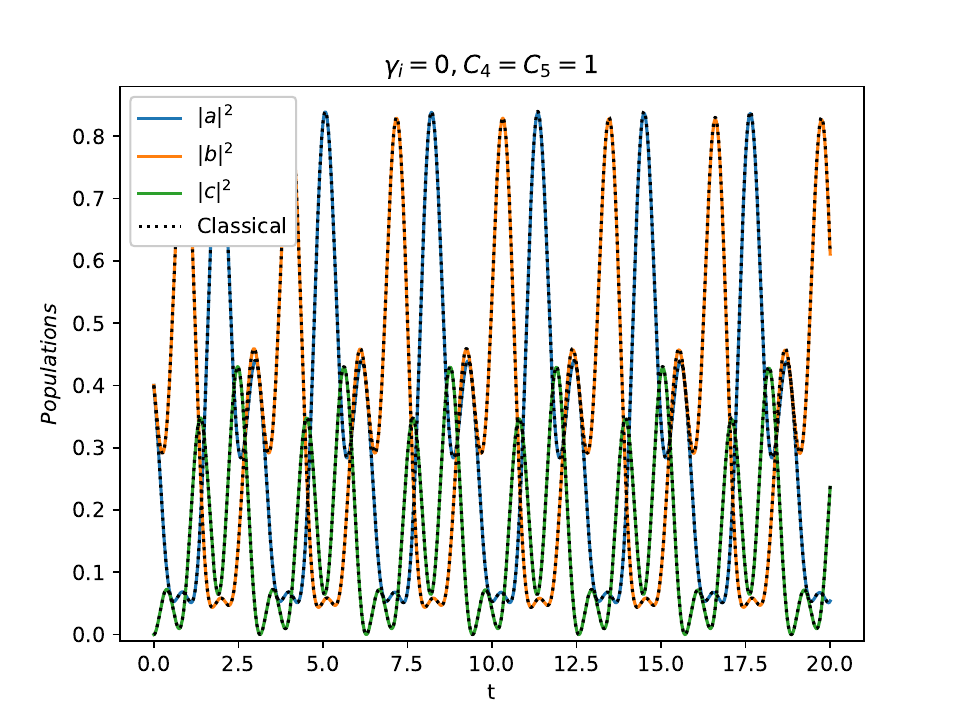}
    \end{subfigure}
    \hspace{0.02\textwidth}
    \begin{subfigure}[b]{0.34\textwidth}
        \includegraphics[width=\textwidth]{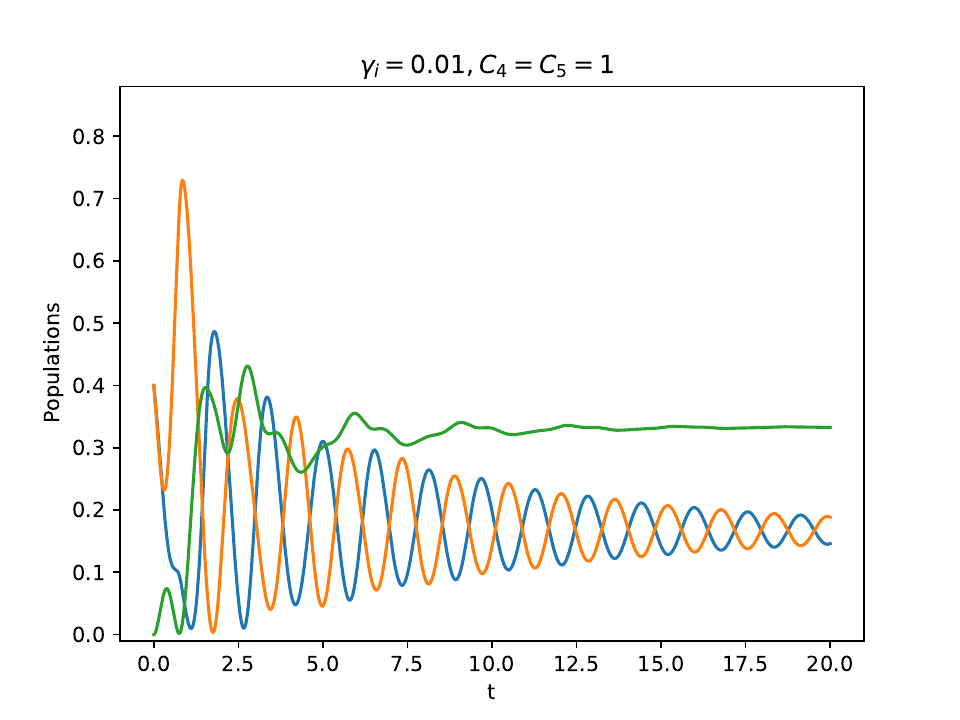}
    \end{subfigure}

    \vspace{0.5em} 

    \begin{subfigure}[b]{0.34\textwidth}
        \includegraphics[width=\textwidth]{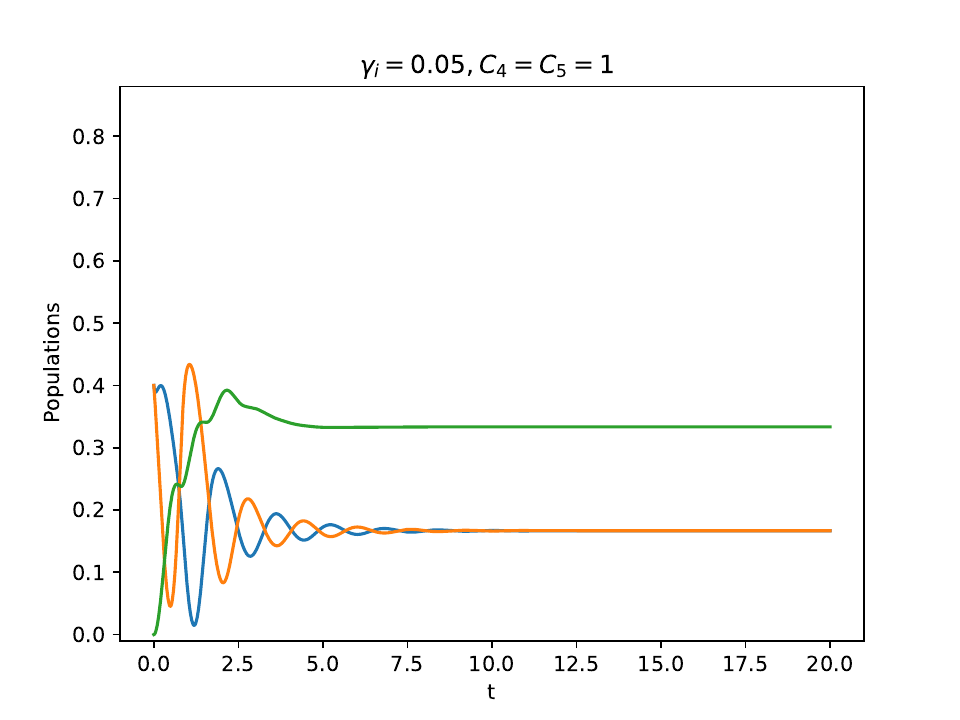}
    \end{subfigure}
    \hspace{0.02\textwidth}
    \begin{subfigure}[b]{0.34\textwidth}
        \includegraphics[width=\textwidth]{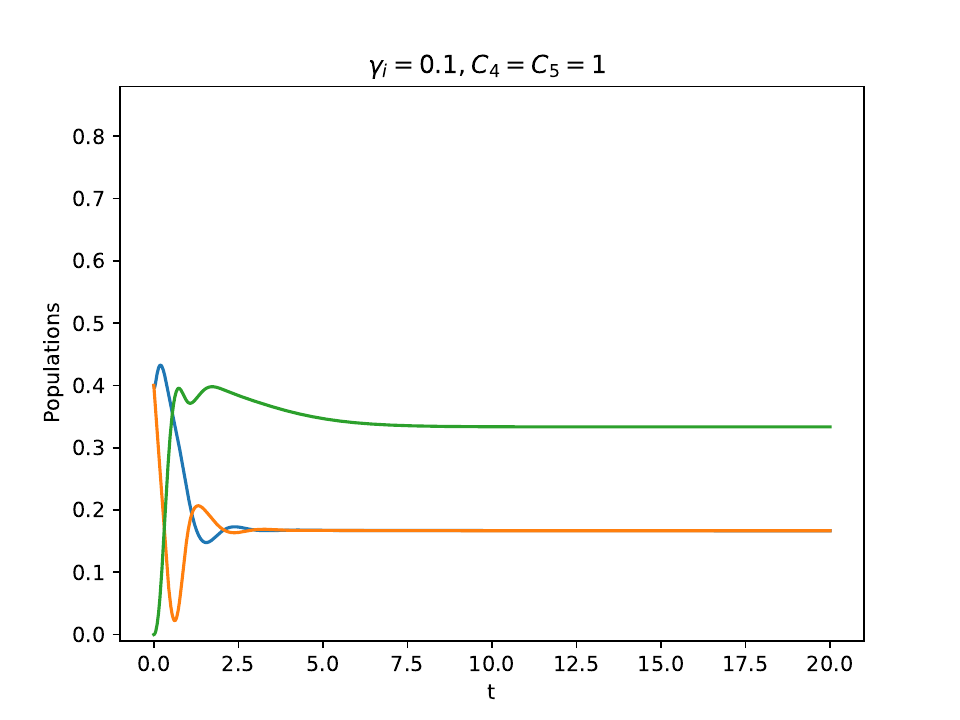}
    \end{subfigure}

    \caption{Time dependent populations varying $\gamma_i$, considering $C_4 = C_5 = 1$.}
    \label{Populations2}
\end{figure}

In contrast, Fig. \ref{Populations2} illustrates the dynamics when the entangling terms of the Hamiltonian are included, with $C_4 = C_5 = 1$. The results exhibit similarities to the non-entangling case, although in Fig. \ref{Populations2} the populations $|b|^2$ and $|c|^2$ converge to a common value over time. This behavior underscores the influence of entanglement, driven by the interaction terms in the Hamiltonian, which encourages correlations between the qubit states. As in the previous figure, the top-left panel displays the exact quantum-classical correspondence for the system, while the remaining panels use the classical framework to describe the dynamics.

Another key quantity for our analysis is the quaternionic population difference, defined as
\begin{equation}
    z(t) \equiv |q_1|^2 - |q_2|^2 = |a|^2 + |b|^2 - |c|^2 - |d|^2,
\end{equation}
where the quaternions $q_1$ and $q_2$ are given by
\begin{align}
    q_1 &= a + b i, & q_2 &= c + d i.
\end{align}

In terms of the coordinates of $\mathbb{CP}^3$, the quaternionic population difference is expressed as
\begin{equation}
    z(t) = \frac{|\mathtt{x}^0|^2 + |\mathtt{x}^1|^2 - |\mathtt{x}^2|^2 - 1}{\mathcal{N}}.
\end{equation}

\begin{figure}[H]
    \centering
    \begin{subfigure}[b]{0.34\textwidth}
        \includegraphics[width=\textwidth]{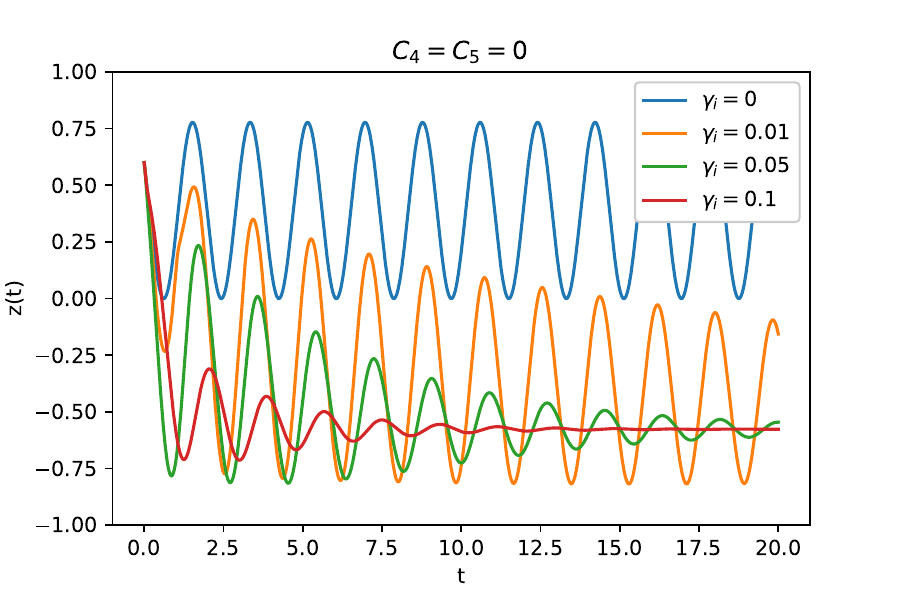}
    \end{subfigure}
    \hspace{0.02\textwidth}
    \begin{subfigure}[b]{0.34\textwidth}
        \includegraphics[width=\textwidth]{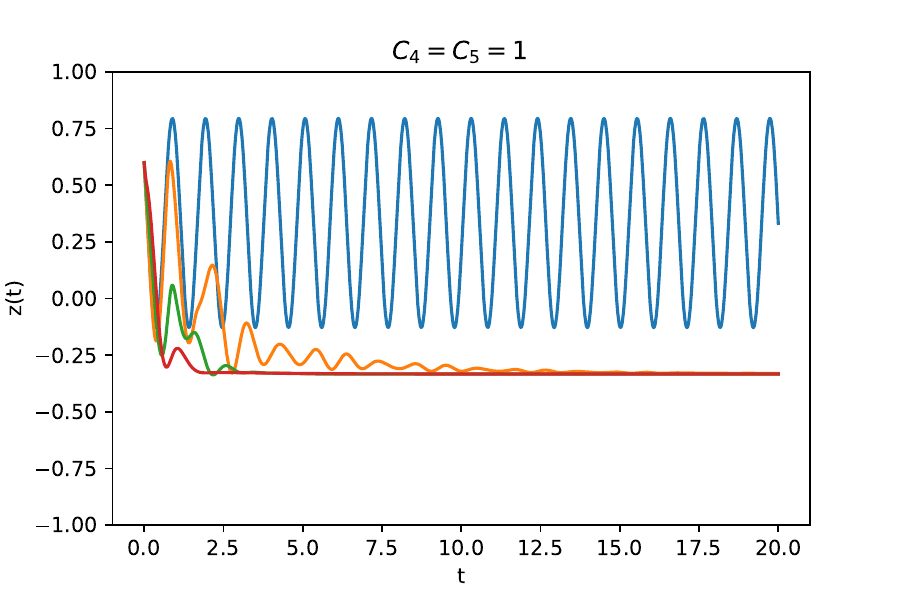}
    \end{subfigure}

    \caption{Time dependent quaternionic population difference varying $\gamma_i$. The entangling terms of the Hamiltonian are only considered in the right panel.}
    \label{populationdifference1}
\end{figure}

As is shown in Fig. \ref{populationdifference1}, the influence of the bath becomes evident as the damping parameter $\gamma_i$ increases.  Specifically, in both panels, the time-averaged quaternionic population difference ($\langle z(t) \rangle_t$) suffers a sign reversal, indicating a significant shift in the two-qubit system’s state probabilities. For the wave function defined as \eqref{wavefunction},
the quantity $z(t) = (|a|^2 + |b|^2) - (|c|^2 + |d|^2)$ represents the difference in probabilities between states where the first qubit is in $\ket{0}$ (i.e., $\ket{00}$ or $\ket{10}$) and those where it is in $\ket{1}$ (i.e., $\ket{01}$ or $\ket{11}$). 
This redistribution of populations, induced by the dissipative interaction,  not only suppresses oscillatory dynamics but also fundamentally alters the quantum state’s statistical distribution, providing an interesting perspective on the behavior of open quantum systems within the $\mathbb{CP}^3$ framework.

Finally, to quantify the entanglement of the two-qubit system, we employ the concurrence, as defined by Wootters~\cite{wootters}. For our system, the concurrence is expressed as $C = 2 |ad - bc|$~\cite{Mosseri_2001}. In the context of our classicalized framework in $\mathbb{CP}^3$, the concurrence can be reformulated in terms of the projective coordinates as
\begin{equation}
    C(t) = 2 |ad - bc| = 2 \frac{|\mathtt{x}^0 - \mathtt{x}^1 \mathtt{x}^2|}{\mathcal{N}}.
\end{equation}

\begin{figure}[H]
    \centering
    \begin{subfigure}[b]{0.34\textwidth}
        \includegraphics[width=\textwidth]{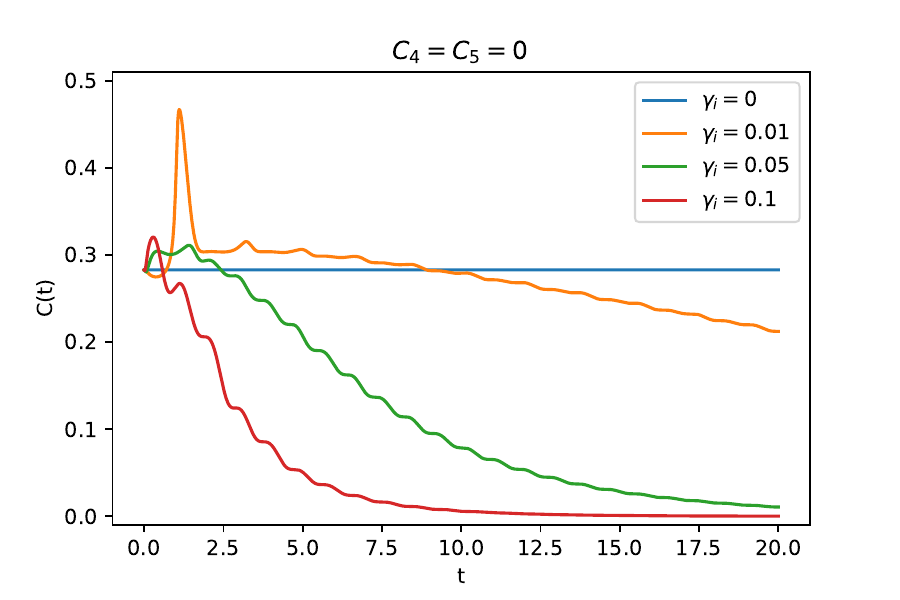}
    \end{subfigure}
    \hspace{0.02\textwidth}
    \begin{subfigure}[b]{0.34\textwidth}
        \includegraphics[width=\textwidth]{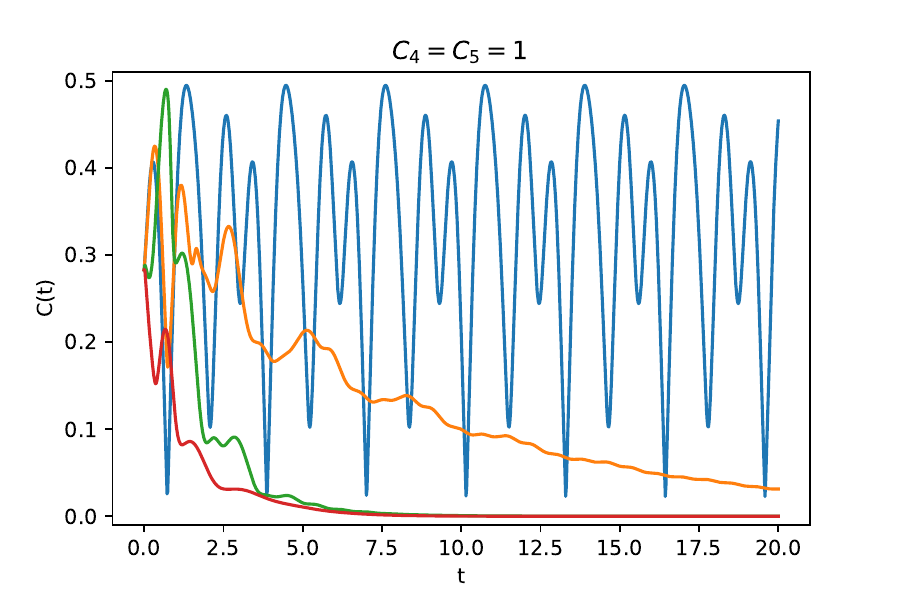}
    \end{subfigure}

    \caption{Time dependent concurrence varying $\gamma_i$. In the left panel, we analyze the dynamics considering the non-entangling terms of the Hamiltonian, while in the right panel, we include the entangling terms.}
    \label{concurrence}
\end{figure}

As is illustrated in Fig. \ref{concurrence}, the concurrence is significantly influenced by the presence of the classical bath.
 In both panels, we observe that, whether we consider cases with or without the entangling terms of the Hamiltonian, increasing environmental damping drives the concurrence to approach zero. This effect indicates a progressive loss of entanglement in the two-qubit system, as the bath’s dissipative effects suppress the quantum correlations captured by $C(t)$. The decreasing concurrence reflects a decoherence effect, effectively driving the system toward a separable state within the $\mathbb{CP}^3$ framework. 
Therefore, under dissipative conditions, the dynamics of the two-qubit system in $\mathbb{CP}^3$ evolve toward separable states, effectively described by the product manifold $\mathbb{CP}^1 \times \mathbb{CP}^1 \approx S^2 \times S^2$, which defines a Segre embedding \cite{book_geometry_quantum}. 

\section{Fenna-Matthews-Olson complex+bath}\label{fmo}

Another significant example of an $N$-state system to which our model can be applied is the Fenna-Matthews-Olson (FMO) complex. This pigment-protein complex, found in green sulfur bacteria, is instrumental in the initial stages of photosynthesis, mediating the transfer of energy \cite{FennaMatthews1975}. In our framework, we represent the FMO complex as a seven-level system, where each diabatic state corresponds to an exciton localized on one of the seven sites. This formalism enables us to explore the energy transfer dynamics within the classicalization method described in this work.

In our previous work, we demonstrated the validity of this model by applying it to a two-qubit entanglement case \cite{dani}. The FMO complex provides an additional case study, reinforcing the exact five-step method to classicalize $N$-state systems.
Furthermore, we will also consider a classical bath of harmonic oscillators, in line with the previous sections, offering a fully classical-like formalism to study dissipative effects in a biological quantum system.

The FMO complex represented as a 7-level quantum system is governed by the following Hamiltonian \cite{AdolphsRenger2006, Richardson2}
\begin{equation}\label{FMO_Ham}
\hat{H}_s =
\begin{pmatrix}
12410 & -87.7 & 5.5 & -5.9 & 6.7 & -13.7 & -9.9 \\
-87.7 & 12530 & 30.8 & 8.2 & 0.7 & 11.8 & 4.3 \\
5.5 & 30.8 & 12210 & -53.5 & -2.2 & -9.6 & 6.0 \\
-5.9 & 8.2 & -53.5 & 12320 & -70.7 & -17.0 & -63.3 \\
6.7 & 0.7 & -2.2 & -70.7 & 12480 & 81.1 & -1.3 \\
-13.7 & 11.8 & -9.6 & -17.0 & 81.1 & 12630 & 39.7 \\
-9.9 & 4.3 & 6.0 & -63.3 & -1.3 & 39.7 & 12440
\end{pmatrix}.
\end{equation}

Following the general procedure described earlier, this system can be classicalized on the complex projective space $\mathbb{CP}^6$. We can apply the same five-step framework in the classicalization process. First of all, the quantum state can be expressed as
\begin{equation}
    \ket{\psi} = a^0\ket{\phi_0}+a^1\ket{\phi_1}+a^2\ket{\phi_2}+a^3\ket{\phi_3}+a^4\ket{\phi_4}+a^5\ket{\phi_5}+a^6\ket{\phi_6},
\end{equation}
so we can obtain the coordinates of $\mathbb{CP}^6$ as
\begin{align}
    x^0 & = \frac{a^0}{a^6},&x^1 & = \frac{a^1}{a^6},&x^2 & = \frac{a^2}{a^6},&x^3 & = \frac{a^3}{a^6},&x^4 & = \frac{a^4}{a^6},&x^5 & = \frac{a^5}{a^6}.
\end{align}

In terms of the $\mathbb{CP}^6$ coordinates, the normalized state becomes
\begin{align}
    \ket{\psi} &= \frac{1}{\sqrt{\mathcal{N}}}(x^0\ket{\phi_0}+x^1\ket{\phi_1}+x^2\ket{\phi_2}+x^3\ket{\phi_3}+x^4\ket{\phi_4}+x^5\ket{\phi_5},
\end{align}
where the normalization factor $\mathcal{N}$ is
\begin{equation}
    \mathcal{N} = 1+\abs{x^0}^2+\abs{x^1}^2+\abs{x^2}^2+\abs{x^3}^2+\abs{x^4}^2+\abs{x^5}^2.
\end{equation}

Now, the classical Hamiltonian function on $\mathbb{CP}^6$ is obtained by computing the expectation value $H_S = \bra{\psi}\hat{H}\ket{\psi}$, where $\hat{H}$ is the Hamiltonian given in Eq. \eqref{FMO_Ham}.

In this case, knowing the inverse symplectic form matrix ($\omega^{j \bar{k}} = -i \mathcal{N} \left( \delta^{j k} + x^j \bar{x}^k \right)$) we obtain 6 Hamilton equations, which can be expressed as
\begin{equation}
\dot{x}^j = \sum_{k=0}^{5} -i \mathcal{N} (\delta^{j k} + x^j \bar{x}^k)\frac{\partial H_S}{\partial \bar{x}^k},
\end{equation}
which only considers the system's classical framework. As in the previous section,  we can incorporate a classical bath of harmonic oscillators. The equations of motion for the noise-averaged system coordinates $\mathtt{x}^j$ in $\mathbb{CP}^6$ are then modified as follows
\begin{equation}
\dot{\mathtt{x}}^j = \sum_{k=0}^{5} -i \mathcal{N} \left( \delta^{jk} + \mathtt{x}^j \bar{\mathtt{x}}^k \right) \left( \frac{\partial H_S}{\partial \bar{\mathtt{x}}^k} + \frac{\partial H_I}{\partial \bar{\mathtt{x}}^k} \right),
\end{equation}
where the interaction term $\frac{\partial H_I}{\partial \bar{\mathtt{x}}^k}$ is given by Eq. \eqref{partial_HI}, as defined earlier. Consequently, the complete system-plus-bath dynamics can be fully simulated using only six differential equations. Therefore, we can now proceed to show the results.
\subsection{Results and discussion}

\begin{figure}[H]
    \centering
    \begin{subfigure}[b]{0.34\textwidth}
        \includegraphics[width=\textwidth]{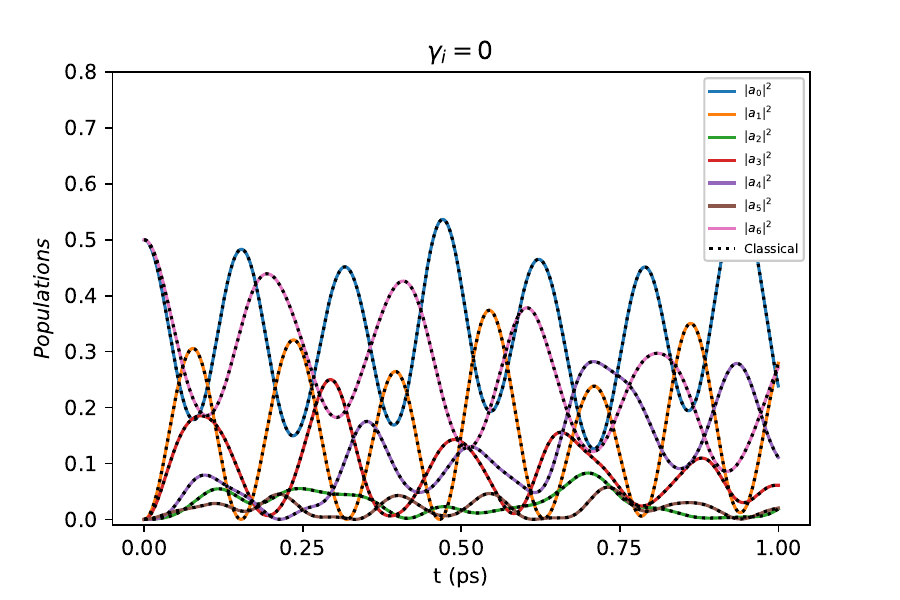}
    \end{subfigure}
    \hspace{0.02\textwidth}
    \begin{subfigure}[b]{0.34\textwidth}
        \includegraphics[width=\textwidth]{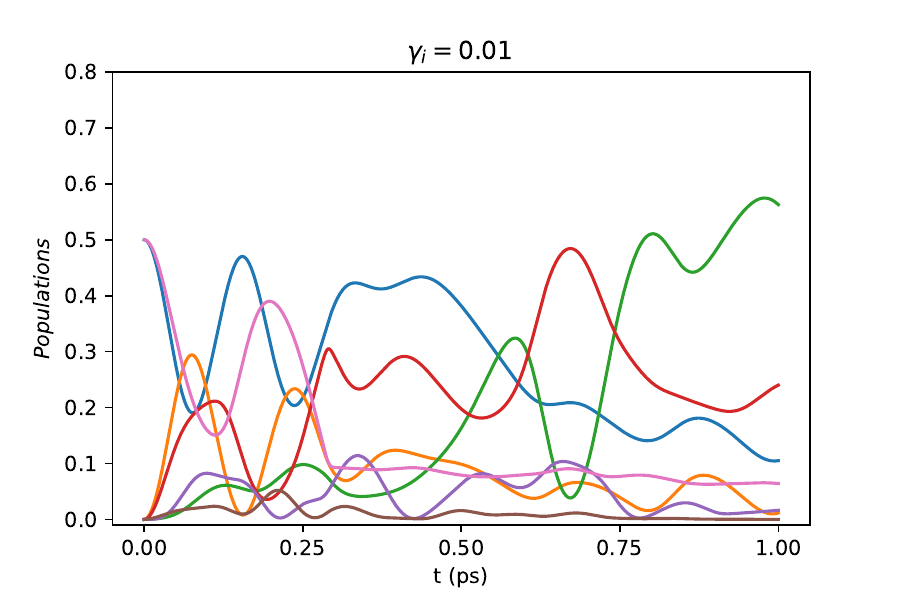}
    \end{subfigure}

    \vspace{0.5em} 

    \begin{subfigure}[b]{0.34\textwidth}
        \includegraphics[width=\textwidth]{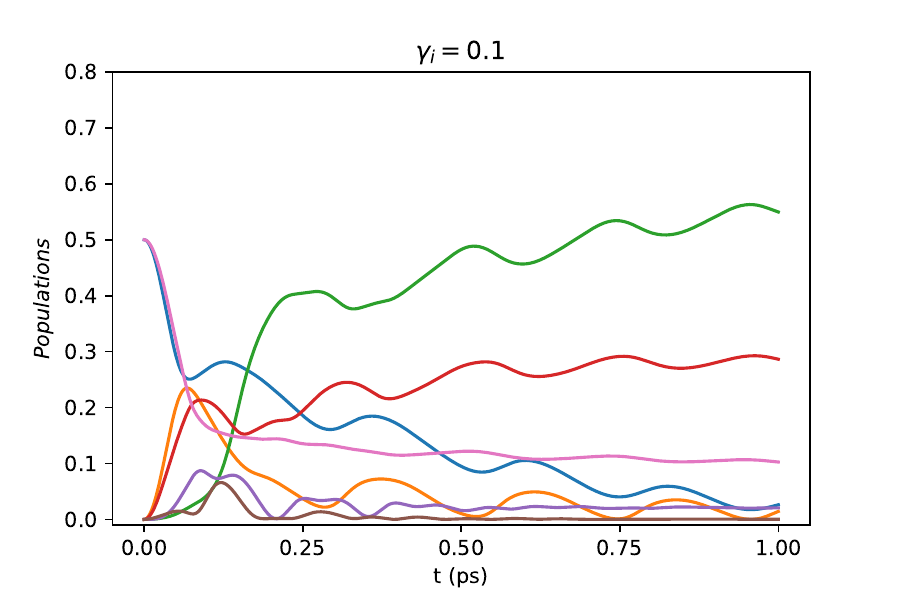}
    \end{subfigure}
    \hspace{0.02\textwidth}
    \begin{subfigure}[b]{0.34\textwidth}
        \includegraphics[width=\textwidth]{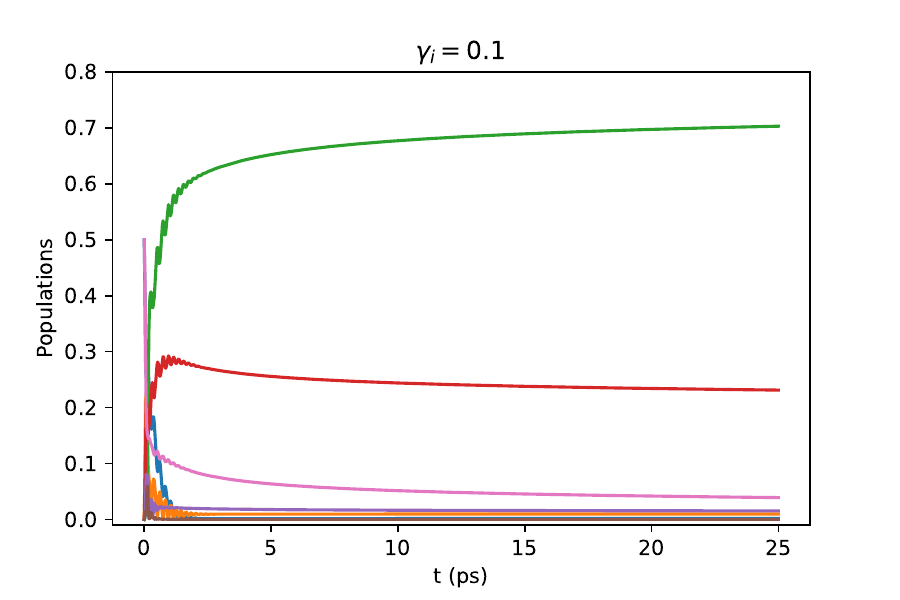}
    \end{subfigure}

    \caption{Time dependent populations changing the value of the damping parameter $\gamma_i$. Top-left panel also shows the classical correspondence and the legend of the present figure.}
    \label{FENNApopulations1}
\end{figure}

In this section, we analyze FMO complex results, which are all compiled in Fig. \ref{FENNApopulations1}.
As is shown, a time scale on the order of picoseconds has been selected. Given that the dynamics of the FMO complex can occur on timescales of femtoseconds to picoseconds, we consider it essential to employ a consistent timescale in our simulations.

The four panels illustrate the time evolution of the populations across the seven sites. Beginning with the top-left panel of Fig. \ref{FENNApopulations1}, we present a comparison between the classical analogue of the isolated FMO complex and its quantum counterpart. The exact correspondence observed serves as further validation of the five-step method's accuracy.

Additionally, Fig. \ref{FENNApopulations1} examines the impact of a harmonic oscillator bath on the FMO complex. Employing the same damping parameters as in the entanglement study, we observe that increasing damping leads to localization of the state populations. Notably, specific populations, such as that of site $a_2$, exhibit a remarkable increase when the bath is incorporated.
Lastly, the bottom-right panel extends the time scale, revealing how damping induces complete stabilization of the states beyond approximately 5 ps.

\section{Conclusions}\label{conclusion}
In this work, we have introduced an exact classicalization method, originally developed for isolated $N$-level quantum systems \cite{dani}, extending it to open quantum systems through coupling with a classical bath of harmonic oscillators. By utilizing the structure of complex projective spaces ($\mathbb{CP}^{N-1}$), a Kähler manifold equipped with a Fubini-Study metric and symplectic form, we transform the quantum evolution of both the system and its bath into a set of $N-1$ Hamilton’s equations through a Langevin formalism, representing a generalized Caldeira-Legget model in a complex projective space. This method maintains the exact quantum behavior of the system, including intricate phenomena like entanglement, while efficiently incorporating the influence of the environment.

To show this extended framework, we applied it to two different systems. First, we have analyzed the dynamics of two entangled qubits, a system previously studied in isolation \cite{dani}, now under the influence of a pure dissipative classical bath. This application demonstrated the method’s ability to capture the impact of dissipation on quantum observables, such as state populations, quaternionic population differences, and concurrence. Second, we have explored the seven-state Fenna-Matthews-Olson (FMO) complex, a biologically relevant system that models energy transfer in photosynthesis. By modeling it, we have shown other example which can be exactly classicalize with the five-step method, also considering the harmonic oscillators' bath.

The strength of this method lies in its generality and computational efficiency. Unlike traditional quantum simulation techniques that often scale exponentially with system size or rely on approximations, the method employed in this manuscript is universal, applicable to any $N$-state quantum system coupled to a harmonic oscillator bath. The geometric foundation of $\mathbb{CP}^{N-1}$ ensures predictable scaling and enables exact simulations of larger systems, such as multi-qubit networks or higher-dimensional qudits. By reducing the dynamics to $N-1$ classical-like equations, even in the presence of environmental interactions, this framework offers a powerful alternative method and a different, more geometric point of view.

In summary, in this work, we have introduced an exact framework for classicalizing $N$-level systems, considering it as an open quantum system, and applying it to two-qubit entanglement dynamics and the FMO complex. This method, although geometrically complex, enables the study of high-dimensional systems under a quantum-classical correspondence in the presence of environmental effects, providing a useful tool for simulations across physics, chemistry, and quantum biology.

\section*{Acknowledgements}
D. M. -G. acknowledges Fundación Humanismo y Ciencia for financial support. D. M. -G. and P. B. acknowledge Generalitat Valenciana through PROMETEO PROJECT CIPROM/2022/13. S. M.-A. acknowledges support of a grant from the Ministry of Science, Innovation and Universities with Ref. PID2023-149406NB-I00.

\bibliographystyle{unsrt}
\bibliography{referencias.bib}

\begin{thebibliography}{10}

\bibitem{QT1}
G.~Benenti, G.~Casati, K.~Saito, and R.~S. Whitney.
\newblock Fundamental aspects of steady-state conversion of heat to work at the nanoscale.
\newblock {\em Physics reports}, 694:1–124, 2017.

\bibitem{QT2}
M.~Carrega, L.~Razzoli, P.~A. Erdman, F.~Cavaliere, G.~Benenti, and M.~Sassetti.
\newblock Dissipation-induced collective advantage of a quantum thermal machine.
\newblock {\em AVS quantum science}, 6(2), 2024.

\bibitem{QT3}
W.~B. Gao, A.~Imamoglu, H.~Bernien, and R.~Hanson.
\newblock Coherent manipulation, measurement and entanglement of individual solid-state spins using optical fields.
\newblock {\em Nature photonics}, 9(6):363–373, 2015.

\bibitem{QT4}
M.~Josefsson, A.~Svilans, A.~M. Burke, E.~A. Hoffmann, S.~Fahlvik, C.~Thelander, M.~Leijnse, and H.~Linke.
\newblock A quantum-dot heat engine operating close to the thermodynamic efficiency limits.
\newblock {\em Nature nanotechnology}, 13(10):920–924, 2018.

\bibitem{QT5}
R.~Kosloff and A.~Levy.
\newblock Quantum heat engines and refrigerators: continuous devices.
\newblock {\em Annual review of physical chemistry}, 65(1):365–393, 2014.

\bibitem{QT6}
A.~Ronzani, B.~Karimi, Y.-C. Senior, J.and~Chang, J.~T. Peltonen, C.~Chen, and J.~P. Pekola.
\newblock Tunable photonic heat transport in a quantum heat valve.
\newblock {\em Nature physics}, 14(10):991–995, 2018.

\bibitem{FennaMatthews1975}
R.~E. Fenna and B.~W. Matthews.
\newblock Chlorophyll arrangement in a bacteriochlorophyll protein from chlorobium limicola.
\newblock {\em Nature}, 258(5536):573--577, 1975.

\bibitem{AdolphsRenger2006}
J.~Adolphs and T.~Renger.
\newblock How proteins trigger excitation energy transfer in the fmo complex: A simple electrostatic model.
\newblock {\em Biophysical Journal}, 91(10):2778--2797, 2006.

\bibitem{strocchi}
F.~Strocchi.
\newblock Complex coordinates and quantum mechanics.
\newblock {\em Rev. Mod. Phys.}, 38:36--40, 1966.

\bibitem{Meyermiller}
H.~D. Meyer and W.~H. Miller.
\newblock A classical analog for electronic degrees of freedom in nonadiabatic collision processes.
\newblock {\em J. Chem. Phys.}, 70:3214–3223, 1979.

\bibitem{StockThoss}
G.~Stock and M.~Thoss.
\newblock Semiclassical description of nonadiabatic quantum dynamics.
\newblock {\em Phys. Rev. Lett.}, 78:578--581, 1997.

\bibitem{Tully1990}
J.~C. Tully.
\newblock Molecular dynamics with electronic transitions.
\newblock {\em J. Chem. Phys.}, 93:1061--1071, 1990.

\bibitem{CottonMiller}
S.~J. Cotton and W.~H. Miller.
\newblock A symmetrical quasi-classical spin-mapping model for the electronic degrees of freedom in non-adiabatic processes.
\newblock {\em The Journal of Physical Chemistry A}, 119:12138–12145, 2015.

\bibitem{Richardson1}
J.~E. Runeson and J.~O. Richardson.
\newblock Spin-mapping approach for nonadiabatic molecular dynamics.
\newblock {\em J. Chem. Phys.}, 151:044119, 2019.

\bibitem{Richardson2}
J.~E. Runeson and J.~O. Richardson.
\newblock Generalized spin mapping for quantum-classical dynamics.
\newblock {\em J. Chem. Phys.}, 152:084110, 2020.

\bibitem{dani}
Daniel Martínez-Gil, Pedro Bargueño, and Salvador Miret-Artés.
\newblock An exact five-step method for classicalizing n-level quantum systems: Application to quantum entanglement dynamics, 2025.

\bibitem{Moroianu_2007}
A.~Moroianu.
\newblock {\em Lectures on Kähler Geometry}.
\newblock London Mathematical Society Student Texts. Cambridge University Press, 2007.

\bibitem{Bellmann}
W.~Ballmann.
\newblock {\em Lectures on Kähler Manifolds}.
\newblock European Mathematical Society, 2006.

\bibitem{Nakahara}
M.~Nakahara.
\newblock {\em Geometry, Topology and Physics}.
\newblock CRC Press, 2003.

\bibitem{chern-1979}
S.-S. Chern.
\newblock {\em {Complex Manifolds without Potential Theory}}.
\newblock Springer New York, NY, 1 1979.

\bibitem{Griffiths}
P.~Griffith and J.~Harris.
\newblock {\em Principles of Algebraic Geometry}.
\newblock Wiley-Interscience, 8 1994.

\bibitem{harris-1992}
Joe Harris.
\newblock {\em {Algebraic geometry}}.
\newblock Springer New York, NY, 1 1992.

\bibitem{fubini}
G.~Fubini.
\newblock Sulle metriche definite da una forma hermitiana.
\newblock {\em Atti Instituto Veneto}, 6:501, 1903.

\bibitem{study}
E.~Study.
\newblock Kürzeste wege in komplexen gebiet.
\newblock {\em Math. Annalen}, 60:321, 1905.

\bibitem{arnold}
V.I. Arnold.
\newblock Symplectic geometry and topology.
\newblock {\em J. Math. Phys.}, 41(6):3307–3343, 2000.

\bibitem{Lee}
J.M.Lee.
\newblock {\em Introduction to Smooth Manifolds}.
\newblock Springer New York, NY, 2012.

\bibitem{Kibble}
Kibble T.W.B.
\newblock Geometrization of quantum mechanics.
\newblock {\em Commun.Math. Phys}, 65:189--201, 1979.

\bibitem{GIBBONS1992147}
G.W. Gibbons.
\newblock Typical states and density matrices.
\newblock {\em Journal of Geometry and Physics}, 8(1):147--162, 1992.

\bibitem{Ashtekar1999}
A.~Ashtekar and T.~A. Schilling.
\newblock {\em Geometrical Formulation of Quantum Mechanics}, pages 23--65.
\newblock Springer New York, New York, NY, 1999.

\bibitem{BRODY200119}
D.~C. Brody and L.~P. Hughston.
\newblock Geometric quantum mechanics.
\newblock {\em Journal of Geometry and Physics}, 38(1):19--53, 2001.

\bibitem{weiss}
U.~Weiss.
\newblock {\em Quantum Dissipative Systems}.
\newblock Series in Modern Condensed Matter Physics, World Scientific, Singapore, 1999.

\bibitem{Schlosshauer2007}
Maximilian Schlosshauer.
\newblock {\em Decoherence and the Quantum–to–Classical Transition}.
\newblock The Frontiers Collection. Springer Berlin Heidelberg, Berlin, Heidelberg, 1 edition, 2007.

\bibitem{book_geometry_quantum}
I.~Bengtsson and K.~Zyczkowski.
\newblock {\em Geometry of Quantum States: An Introduction to Quantum Entanglement}.
\newblock Cambridge University Press, 2006.

\bibitem{caldeiraleggett1}
A.O. Caldeira and A.J. Leggett.
\newblock Path integral approach to quantum brownian motion.
\newblock {\em Physica A: Statistical Mechanics and its Applications}, 121:587--616, 1983.

\bibitem{Caldeiralegget2}
A.~O Caldeira and A.~J. Leggett.
\newblock Influence of dissipation on quantum tunneling in macroscopic systems.
\newblock {\em Phys. Rev. Lett.}, 46:211--214, 1981.

\bibitem{CaldeiraLegget3}
A.O Caldeira and A.J Leggett.
\newblock Quantum tunnelling in a dissipative system.
\newblock {\em Annals of Physics}, 149:374--456, 1983.

\bibitem{RevModPhys.59.1}
A.~J. Leggett, S.~Chakravarty, A.~T. Dorsey, M.~P.~A. Fisher, A.~Garg, and W.~Zwerger.
\newblock Dynamics of the dissipative two-state system.
\newblock {\em Rev. Mod. Phys.}, 59:1--85, 1987.

\bibitem{THORWART2004333}
M.~Thorwart, E.~Paladino, and M.~Grifoni.
\newblock Dynamics of the spin-boson model with a structured environment.
\newblock {\em Chemical Physics}, 296:333--344, 2004.

\bibitem{PhysRevA.68.034301}
T.~A. Costi and R.~H. McKenzie.
\newblock Entanglement between a qubit and the environment in the spin-boson model.
\newblock {\em Phys. Rev. A}, 68:034301, 2003.

\bibitem{PhysRevB.71.035318}
D.~P.~Di Vincenzo and D.~Loss.
\newblock Rigorous born approximation and beyond for the spin-boson model.
\newblock {\em Phys. Rev. B}, 71:035318, 2005.

\bibitem{PedroCPL2011}
P.~Bargueño, H.~C. Peñate-Rodríguez, I.~Gonzalo, F.Sols, and S.~Miret-Artés.
\newblock Friction-induced enhancement in the optical activity of interacting chiral molecules.
\newblock {\em Chemical Physics Letters}, 516:29--34, 2011.

\bibitem{Pedro2012}
A.~Dorta‑Urra, H. C. Peñate‑Rodríguez, P.~Bargueño, G.~Rojas‑Lorenzo, and S.~Miret‑Artés.
\newblock Dissipative geometric phase and decoherence in parity‑violating chiral molecules.
\newblock {\em The Journal of Chemical Physics}, 136(17):174505, 2012.

\bibitem{wootters}
W.K. Wootters.
\newblock Entanglement of formation of an arbitrary state of two qubits.
\newblock {\em Phys. Rev. Lett.}, 80:2245--2248, Mar 1998.

\bibitem{Mosseri_2001}
R.~Mosseri and R.~Dandoloff.
\newblock Geometry of entangled states, bloch spheres and hopf fibrations.
\newblock {\em Journal of Physics A: Mathematical and General}, 34(47):10243, nov 2001.

\end{thebibliography}
\end{document}